\documentclass[showpacs,aps]{revtex4}
\usepackage{amssymb}
\usepackage{amsmath}
\usepackage{graphicx}
\usepackage{lscape}
\usepackage{booktabs}
\begin{document}
%

\title{Study on Fractal Characteristics in the e$^+$e$^-$ Collisions at $\sqrt{s}=250$~GeV}

\author{Zeng Ting-ting, Chen Gang\footnote{Corresponding Author: chengang1@cug.edu.cn}, Dong Zi-jian, She Zhi-lei and N. A. Ragab }

\address{School of Mathematics and Physics, China University of Geosciences, Wuhan
430074, China.}

\begin{abstract}
The fractal characteristics of multi-particle final state are first studied in the e$^+$e$^-$ collision of $\sqrt{s}=250$~GeV, with the MC simulation generator Jetset7.4 and Herwig5.9. The results show that this multi-particle final state has a double-Hurst-exponent fractals characteristics, which means that the resulting NFM obey the scaling law well using both the two sets of Hurst exponents to partition the three dimension phase space isotropically and anisotropically. Their fractal indices and the effective fluctuation strength are also predicted.

\end{abstract}


\maketitle

\section{Introduction}	

It is known that there is large non-linear fluctuation events in the space-time evolution of high energy collisions. Firstly, this large local fluctuation in the pseudo-rapidity density distribution was found\cite{cite1} by the Japanese-American Collaborative Emulsion Experimental group(JACEE) in the observation of nuclear interactions of high energy cosmic rays. Later the so-called ¡°nail¡± events with large rapidity fluctuations were successfully observed by the experimental groups UA5\cite{cite2,cite021} and NA22\cite{cite3}. Then people began to believe that this phenomenon may point to the existence of certain non-linear dynamical fluctuations during the space-time evolution in high energy collisions; in the other words, its intrinsic probability density function has a fractal property. In order to observe this non-linear dynamic fluctuation, Bialas et. al\cite{cite4,cite5} proposed the introduction of normalized factorial moments(NFM) to eliminate statistical fluctuation, and then better to describe the dynamics fluctuation. The fractal behavior of probability density was studied by observing scaling invariance of factorial moments, since then, people have been looking for scaling behavior of factorial moments. Ochs proposed that the fractal behavior of the high-energy hadron-hadron collision in the final phase space should exist in the high-dimensional phase space, and Wu and Liu{\cite{cite6,cite7}} proposed that the fractal of the higher-dimensional phase space is self-affine, rather than self-similar. Later the experimental data confirmed that the final state of the multiplicity produced by the hadron-hadron collisions is self-affine fractal in the high-dimensional phase space{\cite{cite8,cite9}}, while the final state particles in the $e^+e^-$ collision experiment(in the $Z^0$ energy region,  $\sqrt s=91.2$GeV) are self-similar fractal structures\cite{cite10,chenHu2002}.
Recently, the non-linear fluctuation of high-energy proton-proton collisions are researched~\cite{ppc0,ppc1,ppc2}

At present, the Chinese Academy of Sciences is preparing to build a ring-shaped electron-positron collider (CEPC) with the center of mass (c.m.) energy of 250~GeV~\cite{cepc1}. We consider that it is necessary to study the non-linear dynamical fluctuations of the final state multiplicity produced by e$^+$e$^-$ collisions namely the fractal characteristics. In this paper, the MC simulation generator Jetset7.4 and Herwig5.9 are used to generate the e$^+$e$^-$ collisions final state multiplicity events in CEPC energy region, analyze its non-linear fractal characteristics, and study the non-linear dynamic fluctuation properties.

\section{Method of analysis}

We can study the dynamic fluctuation characteristics in high-energy multiparticle production using the scaling behavior of the factorial moments. Normalized factorial moments(NFM) is defined as {\cite{cite15}}£º

\begin{equation}\label{eq1}
F_{q}=\frac{1}{M}{\textstyle\sum_{m=1}^{M}}{\frac{\langle n_m(n_m-1)\cdots(n_m-q+1)\rangle}{\langle n_m\rangle^q}}.
\end{equation} Where, $M$ is the partition number of the phase-space region $\triangle$, $n_m$ is the multiplicity in the $m$-th cell and $q$ is the order of the factorial. It is shown, if the power-law scaling $F_q(M)\sim \delta^{-\phi}$ holds when $\delta \to 0$, then dynamic self-similar fluctuations are presented in multiparticle production. It has been pointed out, however, that the anisotropy of phase space has to be taken into account and that the multiparticle final state in a high energy hadron-hadron collision
may be self-affine rather than self-similar. It has shown that the final state phase space of high-energy multiparticle production appears saturation in low-dimensional case, and self-similar or self-affine fractal is appearing in high-dimension{\cite{cite7}}. The self-similar or self-affine of a dynamical fluctuation can then be characterized by the so-called roughness or Hurst exponents{\cite{cite16}}:

\begin{equation}\label{eq2}
H_{ij}=\frac{\ln\lambda_i}{\ln\lambda_j}=\frac{1+\gamma_j}{1+\gamma_i},\ \
(i,j=a,b,c,\ \ i \neq j),
\end{equation} where $\lambda_a$, $\lambda_b$ and $\lambda_c$ are the shrinkage ratios in three different phase space directions, and $\gamma_a$, $\gamma_b$ and $\gamma_c$ are the saturation exponents obtained by fitting the one-dimensional factorial moments as the following equation,

\begin{equation}\label{eq3}
F_q(M_i)=A_i-B_iM_i^{-\gamma_i},\ \ \ (i=a,b,c).
\end{equation} If Hurst exponents $H_{ij}=1$ or $(\gamma_i=\gamma_j)$, the scaling behavior of the factorial moments may be appeared, the final state multiplicity system corresponding to $i,j$ plane is a self-similar fractal, otherwise is the self-affine fractal. This theory has been confirmed in the Hadron-Hadron collision{\cite{NA221998}} and the e$^+$e$^-$ collision{\cite{chenHu2002}}.

Because Hurst exponents is defined in a plane{\cite{cite17}}, one Hurst exponent only reflects or describes the fractal characteristics of one plane in high-energy multiparticle production. For high-dimensional phase space, such as three-dimensional phase space, its scale properties or fractal characteristics are fully described by three different Hurst exponents.

According to the equation (2), if the dynamical fluctuations of the three-dimensional phase space are isotropic, $H_{ab}=H_{bc}=H_{ca}=1$ or $\gamma_a=\gamma_b=\gamma_c$, the equation has only one solution, i.e. multiparticle production system must be a self-similar fractal. However, if those are anisotropic, $H_{ab}\neq H_{bc}\neq H_{ca}\neq 1$, or $\gamma_a\neq \gamma_b\neq \gamma_c$, then multiparticle production system must be a self-affine fractal. However, since $H_{ab}H_{bc}H_{ca}=1$, the equation (2) may have many group solutions.
 We define the total shrinkage ratios of the 3-dimension phase space as:

\begin{equation}\label{eq4}
\lambda^{aniso}=\lambda_a\lambda_b\lambda_c=\lambda_a^{1+H_{ba}+H_{ca}}=\lambda_a^3.
\end{equation} From the Eq.(2) we can get:
\begin{equation}\label{eq5}
\gamma_a=f(\gamma_a,\gamma_b)=\frac{\gamma_b+\gamma_a+2\gamma_b\gamma_a}{2+\gamma_b+\gamma_a} \ \ (\gamma_b<\gamma_a<\gamma_c\ or \ \gamma_c<\gamma_a<\gamma_b).
\end{equation}
It is clear from Eq.(4) and (5), for the anisotropic of phase space, the scaling characteristics of NFM can be observed when anisotropically partitioning the phase space{\cite{NA221998}} as Hurst exponents. In addition, its scaling characteristics is also observed when the phase space is partitioned isotropically in the case of the total shrinkage rate unchanged as Eq.(4) and conformed to Eq.(5). This total shrinkage rate constant phenomenon is called shrinkage rate of conservation{\cite{Lichen,Lichen1}}. This double-Hurst exponent fractals of the multi-particle final states are confirmed in e$^+$e$^-$ collisions{\cite{JPGchen}}.

\section{The results and discussion}

As mentioned above, we discussed the general criteria of the fractal characteristics of the three-dimensional phase space in the high-energy collision. Then, we will calculate the saturation index of one-dimension NFM in e$^+$e$^-$ collisions at CEPC energy region, and analysis the fractal characteristics of multi-particle final states.

\subsection{One-dimensional NFM analysis}

First, the Jetset7.4 and Herwig5.9 model generators are used to generate 100000 events samples of e$^+$e$^-$ collision at $\sqrt {s}=250$~GeV, then the one-dimensional NFM of the multiparticle final state are calculated according to Eq.(1). Here the rapidity $y$, azimuth angle $\varphi$ and traverse momentum $p_t$ are selected as the phase space variables, where the phase space region are $-5\leq y\leq5,0\leq p_t \leq3$~GeV, $0\le\varphi\leq2\pi$. The distributions of the one-dimensional NFM for the multiparticle final state in e$^+$e$^-$ collisions at $\sqrt {s}=250$~GeV are shown in the Fig.1, as a function of the partition numbers M, $M = 1,2,3,\cdots,40$.

\begin{figure} [pt]
\includegraphics[width=0.75\textwidth]{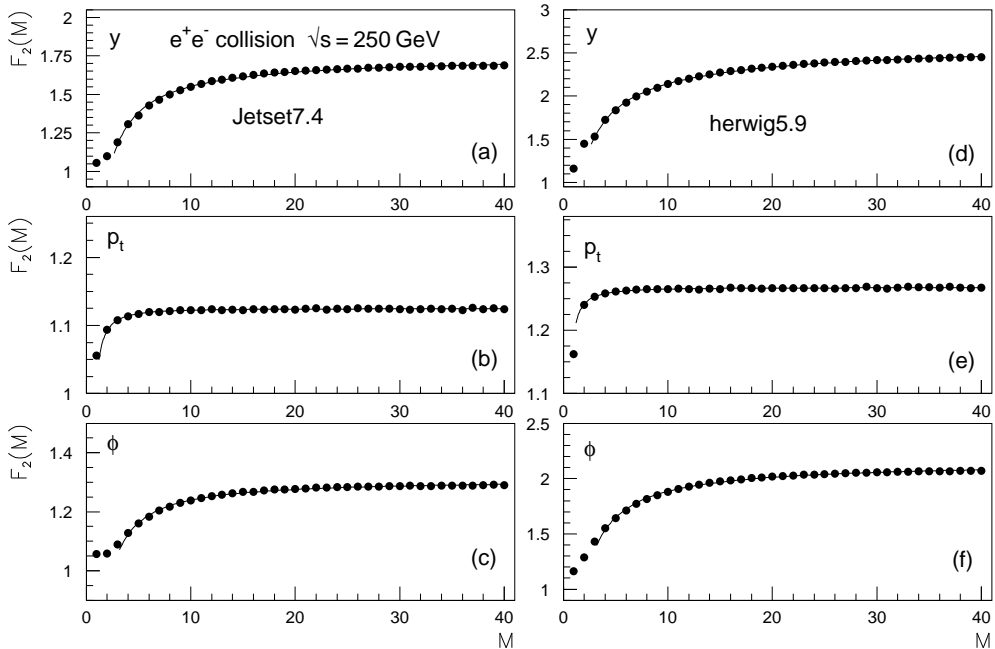}
\caption{Distribution of one-dimensional second-order factorial moments $F_2(M)$ of the multiparticle final state in e$^+$e$^-$ collisions at $\sqrt {s}=250$~GeV, as a function of the partition numbers M. The results by Jetset7.4 on the left $(a) (b) (c)$ are compared with Herwig5.9 on the right $(d) (e) (f)$.
\label{f1}}
\end{figure}

It is clear from Fig.1 that the distributions of one-dimensional second-order NFM are all similar, and tend to be saturated when $M$ increases for the three variables $y,p_t$, and $\varphi$, respectively. The results by Jetset7.4 on the left column of the Fig.1 $(a) (b) (c)$ are similar to that of Herwig5.9 on the right column of the Fig.1 $(d) (e) (f)$.
As W. Ochs pointed out, the fractal is weakened by the superposition effect due to the high-dimensional fractal projection into the low-dimensional phase space, so that the one-dimensional NFM appears saturated when the phase space becomes smaller.

We fit to the data of one-dimensional second-order NFM on the Fig.1 by the Eq.(3), after neglecting the first two points to eliminate the influence of momentum conservation{\cite{liuzy}}. The parameters values for three different variables are listed in Table 1, and the fitted curves are shown as the solid lines in the Fig.1.

\begin{table}[tbp]
\caption{The  parameters values obtained from a fit of one-dimensional second-order NFM by Eq.(3) in e$^+$e$^-$ collisions at $\sqrt{s}=250$~GeV.}
{\begin{tabular}{@{}cccccc@{}} \toprule
Model &Variable & A & B & $\gamma$ & $\chi^2/DF$\\\colrule
 \quad &$\mbox{y} $&$_{1.758\pm 0.002}$&$_{1.50\pm 0.01}$&$_{0.86\pm 0.01}$&$_{1.42}$ \\
{Jetset7.4}&${p_t}$&$_{1.125\pm 0.001}$&$_{0.10\pm 0.05}$&$_{1.61\pm 0.13}$&$_{0.49}$ \\
 \quad &$\varphi  $&$_{1.308\pm 0.002}$&$_{0.83\pm 0.04}$&$_{1.08\pm 0.04}$&$_{0.68}$ \\ \colrule
 \quad &$\mbox{y} $&$_{2.690\pm 0.010}$&$_{2.311\pm 0.002}$&$_{0.62\pm 0.05}$&$_{1.15}$ \\
{Herwig5.9}&${p_t}$&$_{1.268\pm 0.010}$&$_{0.070\pm 0.050}$&$_{1.47\pm 0.23}$&$_{0.98}$ \\
 \quad &$\varphi  $&$_{2.159\pm 0.005}$&$_{2.16\pm 0.05}$&$_{0.90\pm 0.08}$&$_{1.55}$ \\ \botrule
\end{tabular} \label{ta1}}
\end{table}

 According to the Eq.(2), the Hurst exponents $H_{yp_{t}}$, $H_{p_{t}\varphi}$, and $H_{\varphi y}$ are calculated using the saturation exponents in Table 1, as shown in Table 2. The results show that the Hurst exponents of three variables in two models all do not equal to 1, $H_{ab}\neq H_{bc}\neq H_{ca}\neq 1$, or $\gamma_a\neq \gamma_b\neq \gamma_c$ , so the dynamical fluctuations of the final state particle system are anisotropic, and the corresponding fractal structure may be self-affine. Therefore, if the phase space is anisotropic partitioned according to the Hurst exponent, we should observe the scale characteristics of three-dimension NFM.

However, since the Hurst exponent satisfies the equation (5), i.e. $\gamma_\varphi= f(\gamma_\varphi,\gamma_{p_{t}})$, the total shrinkages of isotropic and anisotropic partitioning phase space as the Hurst exponent in Table 2 are equal. So the double Hurst exponent phenomenon exists in the fractal of the final state multiparticle of the e$^+$e$^-$ collisions at $\sqrt s=250$~GeV, i.e. the scaling properties of the three-dimensional NFM are also observed when phase space is  partitioned isotropically.

\begin{table}[pt]
\caption{The saturation exponents are as in table 1 and Hurst exponent are calculated with the saturation exponents by Eq.(4).}
{\begin{tabular}{@{}cccccccc@{}} \toprule
Model & $\gamma_y$ & $\gamma_{p_t}$ & $\gamma_{\varphi}$ & $H_{yp_t}$ & $H_{p_t\varphi}$ & $H_{\varphi y}$\\ \colrule
{Jetset7.4}&$_{0.86\pm 0.01}$&$_{1.61\pm 0.13}$&$_{1.08\pm 0.04}$ &$_{1.40\pm 0.07}$ &$_{0.79\pm 0.05}$ &$_{0.91\pm 0.03}$  \\
{Herwig5.9}&$_{0.62\pm 0.05}$&$_{1.47\pm 0.23}$&$_{0.90\pm 0.08}$ &$_{1.52\pm 0.16}$ &$_{0.77\pm 0.09}$ &$_{0.85\pm 0.06}$  \\ \botrule
\end{tabular} \label{ta2}}
\end{table}

\subsection{Three-Dimensional NFM analysis of anisotropic}

We can perform a self-affine analysis in three-dimensional phase space with the Hurst exponents in Table 2. Since $H_{yp_t}\neq H_{p_t\varphi}\neq H_{\varphi y}\neq 1$, or $\gamma_y\neq \gamma_{p_t}\neq \gamma_\varphi$, from Eq.(2), it follows as
\begin{equation}
M_j=M_i^{\frac{1}{H_{ij}}},\ \ i,j=y,p_t,\varphi ,
 \end{equation} we use a partitioning $M=M_yM_{p_t}M_{\varphi}$, where $M_y= 1,2,3,4,5,\dots,40$; $M_{p_t}= 1,1,2,2,3,\dots,13$, $M_{\varphi}=1,2,3,4,4\dots,27$.
 The results for ln$F_q$ vs. ln$M$ in three-dimensional phase space for orders $q=2-5$ of e$^+$e$^-$ collisions at $\sqrt s=250$~GeV are shown in Fig.2.
 Although the non-integer partition phase space lost a end of the small, its distribution has produced local fluctuations, but $\ln F_q \sim$ ln $M$ distribution continues to show good scaling properties{\cite{cite17,citeChenLG}}. It is shown that the fractal structure is self-affine in the e$^+$e$^-$ collisions at $\sqrt s=250$~~GeV.

In order to show the quality of the scaling law, a linear fits to three-dimensions NFM are compared with the data in Fig. 2 according to the following formula:
\vspace{-3mm}
\begin{equation}\label{eq7}
\ln F_q=c+\phi_q\ln M
\end{equation}
the fitting results are listed in Table 3. To reduce the influence of momentum conservation, the first point are excluded in all the fits.
 It can be seen from Fig.2 that the results give a linear property after the first point for orders $q=2-5$ are omitted. It is shown that the fractal of  multiplicity final state in the e$^+$e$^-$ collisions at $\sqrt s=250$~GeV is self-affine. In addition, we also can see that the results of the MC Jetset7.4 are consistent with the Herwig5.9.

\begin{figure}[pb]
\includegraphics[width=0.75\textwidth]{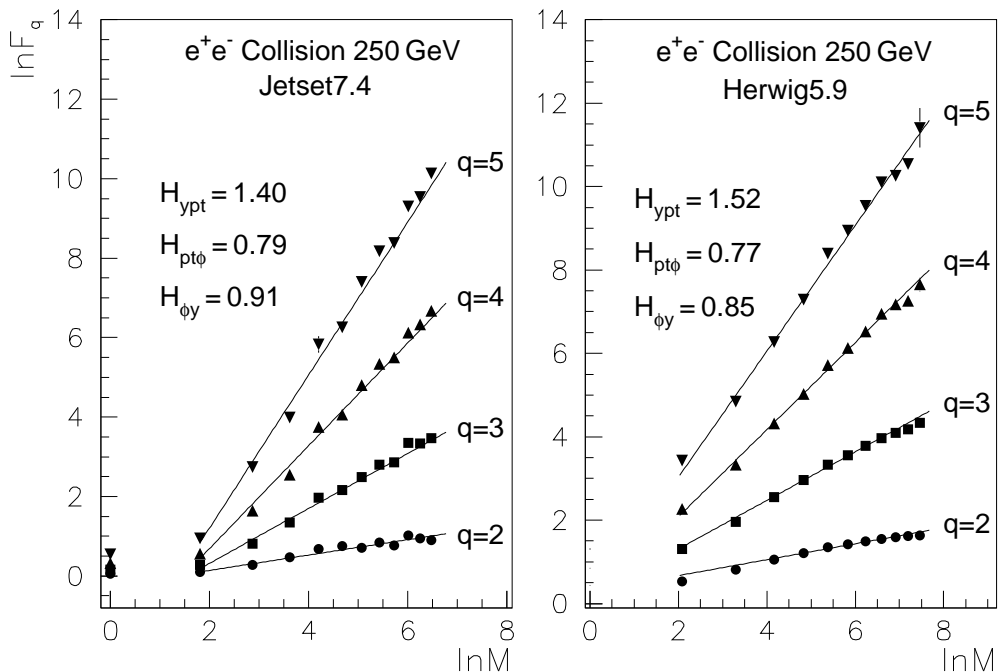}
\caption{The logarithmic distribution of three-dimensional $F_q(M)$ self-affine analysis in e$^+$e$^-$ collisions at $\sqrt {s}=250$~GeV for the order $q=2-5$, as a function of the partition numbers ln$M$. The results by Jetset7.4 on the left is compared with Herwig5.9 on the right. The curves are fitted by Eq.(5).
\label{f2}}
\end{figure}
\begin{table}[pt]
\caption{The parameters values obtained from a fit of the three-dimensional NFM with the self-affine analysis by Eq.(7) in e$^+$e$^-$ collisions at $\sqrt{s}=250$ ~GeV.}
{\begin{tabular}{@{}ccccc@{}} \toprule
Model&q & c & $\phi_q$  & $\chi^2/DF$\\ \colrule
 \quad &${2}$&${-0.07\pm 0.03}$&${0.160\pm 0.006}$&${10.4}$ \\
{Jetset7.4} &${3}$&${-1.4\pm 0.1}$&${0.75\pm 0.03}$&${0.53}$ \\
\quad &${4}$ &${-2.7\pm 0.3}$&${1.46\pm 0.07}$&${0.15}$ \\
\quad &${5}$&${-3.8\pm 0.9}$&${2.2\pm 0.2}$&${0.06}$ \\ \colrule
 \quad &${2}$&${0.288\pm 0.004}$&${0.191\pm 0.001}$&${49.27}$ \\
{Herwig5.9}&${3}$&${0.15\pm 0.02}$&${0.582\pm 0.004}$&${8.90}$ \\
\quad &${4}$&${-0.007\pm 0.001}$&${1.044\pm 0.010}$&${4.57}$ \\
\quad &${5}$&${-0.013\pm 0.001}$&${1.50\pm 0.02}$&${3.05}$ \\ \botrule
\end{tabular} \label{ta3}}
\end{table}

\subsection{Three-dimensional NFM analysis of isotropic}

Although the Hurst exponents of $H_{yp_{t}}$, $H_{p_{t}\varphi}$, and $H_{\varphi y}$ are not equal, their three one-dimensional saturation exponents in Table 2 satisfy the Eq.(4) and (5), i.e. the total shrinkage rate is constant and $\gamma_\varphi=f(\gamma_\varphi,\gamma_{p_{t}})$ within the error range in e$^+$e$^-$ collisions at $\sqrt s=250$~GeV. Thus the multiparticle final states of e$^+$e$^-$ collisions at $\sqrt s=250$~GeV should be able to show the double-Hurst exponent fractals phenomenon. If the phase space is partitioned isotropically, this scale of self-similar of three-dimensional NFM should also be observed{\cite{Lichen}}.

We calculate the three-dimension NFM of e$^+$e$^-$ collisions at $\sqrt s=250$~GeV for order $q=2-5$ using MC model Jetset7.4 and Herwig5.9, where the division numbers on the three-dimensional phase space are selected as $M_\varphi =M_{p_{t}}=M_y = 1,2,3,\cdots,40$.  The results of logarithmic distribution of three-dimensional $\ln F_q(M)$ versus $\ln M$, with the isotropic partition phase space, are plotted in Fig 3. In order to show the quality of the scaling law, a linear fits to three-dimensions NFM are compared with the data in Fig.3 according to Eq.(7). The parameters values are listed in Table 4, omitting the first point to eliminate the influence of momentum conservation in these fits.
We compare the results of the MC Jetset7.4 on the left in Fig.3 with Herwig5.9 on the right in Fig.3, and find that they are consistent.
\begin{figure}[pb]
\includegraphics[width=0.75\textwidth]{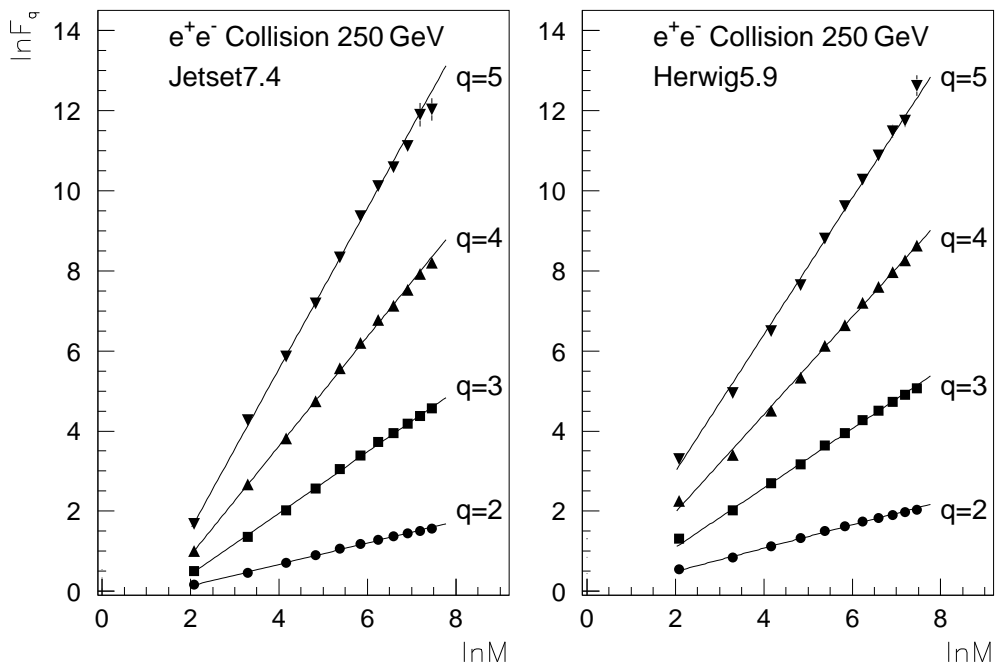}

\caption{The logarithmic distribution of three-dimensional $\ln F_q(M)$ versus $\ln M$ with the isotropic partition phase space, in e$^+$e$^-$ collisions at $\sqrt {s}=250$~GeV for the orders $q=2\sim 5$. The results by Jetset7.4 on the left is compared with Herwig5.9 on the right. The curves are fitted by Eq.(7)}
\end{figure}

From Fig.3, one can see that the distribution of $\ln F_q \sim \ln M$ presents a good linear characteristic. It is shown that the scaling law of NFM for multiplicity final state also can be observed in the case of isotropic partition phase space, if the shrinkage rate of three-dimension phase space can satisfy conditions as Eq.(4) and (5). Since the total shrinkage ratios are the same, that is, the total shrinkage rate is conserved. This is also shown that there is also a self-similar fractal characteristics in multi-particle final states of e$^+$e$^-$ collisions at CEPC energy region.

\begin{table}[pt]
\caption{The parameters values obtained from a fit of the three-dimensional NFM with the isotropic partition phase space in e$^+$e$^-$ collisions at $\sqrt{s}=250$~GeV.}
{\begin{tabular}{@{}ccccc@{}} \toprule
Model&q &C & $\phi_q$  & $\chi^2/DF$\\ \colrule
 \quad &${2}$&${-0.405\pm 0.003}$&${0.268\pm 0.007}$&${18.99}$ \\
{Jetset7.4} &${3}$&${-1.116\pm 0.008}$&${0.766\pm 0.002}$&${10.83}$ \\
\quad &${4}$ &${-1.83\pm 0.02}$&${1.365\pm 0.005}$&${1.85}$ \\
\quad &${5}$&$_{-2.46\pm 0.04}$&${2.00\pm 0.01}$&${1.77}$ \\ \colrule
 \quad &${2}$&${-0.086\pm 0.004}$&${0.290\pm 0.007}$&${36.152}$ \\
{Herwig5.9}&${3}$&${-0.39\pm 0.02}$&${0.741\pm 0.003}$&${5.01}$ \\
\quad &${4}$&${-0.76\pm 0.04}$&${1.269\pm 0.009}$&${1.55}$ \\
\quad &${5}$&${-1.02\pm 0.01}$&${1.80\pm 0.09}$&${0.77}$ \\ \botrule
\end{tabular} \label{ta4}}
\end{table}

\subsection{Effective fluctuation intensity}

In order to further describe this non-linear dynamics fluctuation of the multiplicity final state in the high-energy collisions,
we can also consider that an effective fluctuation strength can be taken as a characteristic quantity for the intensity of dynamical fluctuations, which is defined as follows{\cite{cite21}}:

\begin{equation}\label{eq11}
\alpha_{eff} = \sqrt{{6\ln2 \over q}(1-D_q)}=\sqrt{{6\ln 2\over q}{\phi_q\over {q-1}}},
\end{equation}
where $q$ is the order of the factorial moments, $\phi_q$ is the intermittency index. The effective fluctuation strength $\alpha_{eff}$ in the  e$^+$e$^-$ collisions at $\sqrt s=$ 250~GeV is calculated and filled into the table 5. As a comparison, table 5 also list the effective fluctuation strengths in Au-Au collisions at $\sqrt{s}=200$~GeV, the NA22's  hadron-hadron collisions experiment at $250$~GeV, and the L3's  e$^+$e$^-$ collision experiment at $91.2$~GeV, respectively. Here the intermittency exponents calculated these effective fluctuation strengths are taken from references\cite{chenHu2002,NA221998,NPAxie}.

\begin{table}[ph]
\caption{Comparison of effective fluctuation strengths $\alpha_{eff}$ for hadron-hadron, Au-Au and e$^+$e$^-$ collisions.}
{\begin{tabular}{@{}cccccc@{}} \toprule
$ Model$&AMPT\cite{NPAxie} & NA22\cite{NA221998}& Jetset7.4\cite{chenHu2002} &{Jetset7.4} & {Herwig5.9}\\
$ $&{(200~GeV)} & {(250~GeV)} & {(91.2~GeV)} &{(250~GeV)} & {(250~GeV)}\\ \colrule
q &Au-Au & $\pi^+ (\kappa^+ )p $ & e$^+$e$^-$  & e$^+$e$^-$ & e$^+$e$^-$\\ \colrule
${2}$&${0.025\pm 0.004}$&${0.356\pm 0.011}$&${0.635\pm 0.010}$&${0.747\pm 0.011}$&${0.763\pm 0.012}$ \\
 ${3}$&${0.024\pm 0.001}$&${0.408\pm 0.011}$&${0.644\pm 0.013}$&${0.728\pm 0.002}$&${0.718\pm 0.009}$ \\
 ${4}$&${0.024\pm 0.001}$&${0.496\pm 0.012}$&${0.612\pm 0.009}$&${0.687\pm 0.012}$&${0.664\pm 0.017}$ \\
 ${5}$&${0.024\pm 0.001}$&${0.572\pm 0.017}$&${0.600\pm 0.008}$&${0.645\pm 0.011}$&${0.612\pm 0.001}$ \\ \botrule
\end{tabular} \label{ta4}}
\end{table}

 We can see from Table 5 that the effective fluctuation strengths are approximately constant when the intermittent exponents increase with the order $q$ increasing, and the effective fluctuation strength in e$^+$e$^-$ collisions are much larger than those in heavy ion Au-Au collisions and hadron-hadron collisions. This is a noteworthy phenomenon that it is generally believed that the effective fluctuation strength in heavy ion collisions energy should be larger than that in e$^+$e$^-$ collisions, since it's multiplicity involved is more, the process is more complicated. It would be a sign of the QGP generated. In addition, the effective fluctuation strengths are a slight increase as the collision energy increase
 in e$^+$e$^-$ collisions.

\section{Summary }

\label{sec} In this paper, we introduce an analysis method of the factorial moment to study the dynamical fluctuation properties of the multiplicity production in e$^+$e$^-$ collisions at $\sqrt s=250$~GeV. The MC simulation generator Jetset7.4 and Herwig5.9 are used to generate multi-particle final state events of CEPC energy region. By fitting one-dimensional projection of the factorial moments, the Hurst exponents are derived for all combinations of the phase-space variables as $(y, p_t,\varphi)$, which are not equal,i.e. $H_{ab}\neq H_{bc}\neq H_{ca} \neq 1$. But their total shrinkage rate, satisfied with the Eq.(4) and (5), is conserved. So the multiparticle final states of e$^+$e$^-$ collisions at $\sqrt s=250$~GeV exist double-Hurst-exponent fractals phenomenon, i.e, both partition phase space isotropic or/and anisotropic, the NFM's scaling behavior can be all observed.

It is worth noting that the three-dimension isotropic analysis show a better straight line than the three-dimension anisotropic analysis, or NFM's scaling behavior than those anisotropic one. This is because
partition number $M$ is non-integer as anisotropic analysis, $M=N+a$~\cite{citeChenLG,NA221998}, and the small phase space $a$ will be lost. That will inevitably produce errors or/and increase the fluctuation for the NFM.

In a word, it is found that the multi-particle final states in e$^+$e$^-$ collisions at $\sqrt s=250$~GeV are double-Hurst exponent fractals, which not only shows a self-affine fractals characteristics as Hurst exponent, but also with a self-similar fractals, i.e, it has a multifractal characteristics.

 We also have calculated the effective fluctuation strengths $\alpha_{eff}$ of describing the fluctuation or intermittent intensity in a high-energy collisions. It is found that the effective fluctuation strengths are approximately constant when the intermittent exponents increase with the order $q$ increasing, and the effective fluctuation strengths in e$^+$e$^-$ collisions are much larger than those in heavy ion Au-Au collisions and hadron-hadron collisions.

 The Chinese Academy of Sciences is preparing to build a ring-shaped electron-positron collider (CEPC) with the c.m. energy of $250$~GeV. It is interesting to examine these fractal features in coming CEPC experiments.

\section*{Acknowledgments}
Finally, we acknowledge the financial support from NSFC (11475149). The authors thank PH.D. Li Di-kai for helpful discussions.



\end{document}